\documentclass[12pt]{article}
\date{}
\begin{document}
\title{Baryon flow at SIS energies 
\thanks{Work supported by BMBF and GSI Darmstadt.}}
\author{P. K. Sahu\thanks{Alexander von Humboldt Research Fellow.}, 
A. Hombach, W. Cassing and U. Mosel \\ 
Institut f\"ur Theoretische Physik\\
Universit\"at Giessen\\
D-35392 Giessen\\
GERMANY}
\maketitle

\begin{abstract}

We calculate the baryon flow $\langle P_x/A(y) \rangle$ in the energy 
range from .25 to $\le$2.5$AGeV$
in a relativistic transport model for $Ni+Ni$ and $Au+Au$ collisions employing
various models for the baryon self energies. We find that to describe the flow 
data of the FOPI Collaboration the strength of the vector potential has to be 
reduced at high relative momentum or at high density such that the 
Schr\"odinger- equivalent potential at normal nuclear density decreases 
above 1$GeV$ relative  kinetic energy and approaches zero above 2$GeV$.

\end{abstract}



\newpage
\section{Introduction}
The nuclear equation of state (EOS) at high density ($\rho \ge 3 \rho_0$) 
is still an unsettled issue though many experimental efforts have been made in
the last couple of years to address this question in a more systematic way 
\cite{gut89,par95,her96,cha97}.
Experimentally, the baryon sidewards flow and subthreshold particle 
production are the most promising observables. Recently \cite{her96,rei97}, 
the flow has been measured for heavy-ion collisions at SIS energies 
($\le 2 AGeV$)
for $Ni+Ni$ and $Au+Au$ systems, which provides further constraints on the
hadronic models and the  EOS employed. 

From the theoretical point of view relativistic transport models 
\cite{mar94,ehe96,li97} 
have been used to describe heavy-ion collisions at energies from 
the SIS at GSI to the AGS at BNL and the SPS regime at CERN ($\le 200 AGeV$).
There are two main
ingredients in the transport model: the mean fields (i.e. scalar and vector 
self energies) and in-medium baryon-baryon and meson-baryon cross sections. 
The scalar and vector mean fields are usually derived from some hadronic
Lagrangian (which gives a well defined EOS) while the baryon-baryon and
meson-baryon cross sections are taken from experimental data for the
related processes in free space. In-medium modifications of the inelastic
channels, furthermore, are constrained by the experimental data on the 
particle rapidity distributions which control the amount of 'stopping'.

In principle, the baryon selfenergies should be determined by a 
Dirac-Brueckner approach including all relevant hadronic couplings.
However, calculations for configuration dependent phase-space distributions
are not very reliable yet and limited in density as well as in momentum
\cite{ser86,cla83,bot90,seh90}.

As mentioned before, the scalar and vector self energies for nucleons with 
their
explicit momentum and density dependence are the key quantities that 
determine the nuclear EOS. In this work we will perform a systematic study
of $Ni + Ni$ and $Au + Au$ collisions in order to extract further 
information on these quantities in comparison to the recent experimental
data on the collective flow of baryons.

Our work is organized as follows: In Section 2 we will briefly describe
the relativistic transport approach employed as well as the known constraints
on the momentum dependence of the scalar and vector self energies. Section
3 is devoted to a systematic comparison of the calculated flow - employing
various Lagrangian models - to the experimental data while Section 4
concludes with a summary and discussion of open problems.  

\pagebreak

\section{The transport model}

In this work we perform the theoretical analysis along the line of a
relativistic transport approach
which is based on a coupled set of covariant transport
equations for the phase-space distributions $f_{h} (x,p)$ of a hadron $h$
\cite{ehe96,web93}, i.e.
\begin{eqnarray}  \label{g24}
\lefteqn{\left\{ \left( \Pi_{\mu}-\Pi_{\nu}\partial_{\mu}^p U_{h}^{\nu}
-M_{h}^*\partial^p_{\mu} U_{h}^{S} \right)\partial_x^{\mu}
+ \left( \Pi_{\nu} \partial^x_{\mu} U^{\nu}_{h}+
M^*_{h} \partial^x_{\mu}U^{S}_{h}\right) \partial^{\mu}_p
\right\} f_{h}(x,p) } \nonumber \\
& = &\sum_{h_2 h_3 h_4} \int d2 d3 d4 \,\,
 [G^{\dagger}G]_{12\to 34}\,\,
\delta^4_\Gamma(\Pi +\Pi_2-\Pi_3-\Pi_4  )  \nonumber\\
& \times &\left\{ f_{h_3}(x,p_3)f_{h_4}(x,p_4)\bar{f}_{h}(x,p)
\bar{f}_{h_2}(x,p_2)\right.  \nonumber\\
& - & \left. f_{h}(x,p)f_{h_2}(x,p_2)\bar{f}_{h_3}(x,p_3)
\bar{f}_{h_4}(x,p_4) \right\} \ .
\end{eqnarray}
In Eq.~(\ref{g24}) $U_{h}^{S}(x,p)$ and $U_{h}^{\mu}(x,p)$ denote the
real part of the scalar and vector hadron self energies, respectively,
while $[G^+G]_{12\to 34} \delta^4_\Gamma (\Pi
+\Pi_2-\Pi_3-\Pi_4 )$ is the `transition rate' for the process
$1+2\to 3+4$. Though in quantum many-body systems the transition rate 
is partly off-shell - as indicated by the index $\Gamma$ 
of the $\delta$-function - we use the semi-classical
on-shell limit $\Gamma \rightarrow 0$ since this approximation
is found to describe reasonably well hadronic spectra in a wide 
dynamical regime. The hadron quasi-particle properties in
(\ref{g24}) are defined via the mass-shell constraint 
$\delta (\Pi_{\mu}\Pi^{\mu}-M_{h}^{*2})$ \cite{web93}
with effective masses and momenta given by
\begin{eqnarray} \label{g26}
M_{h}^* (x,p)&=&M_h + U_h^{{S}}(x,p) \nonumber \\
\Pi^{\mu} (x,p)&=&p^{\mu}-U^{\mu}_h (x,p)\ \ ,
\end{eqnarray}
while the phase-space factors
\begin{equation}
\bar{f}_{h} (x,p)=1 - f_{{h}} (x,p)
\end{equation}
account for fermion Pauli-blocking.
The transport approach (\ref{g24}) is fully specified by
$U_{h}^{S}(x,p)$ and $U_{h}^{\mu}(x,p)$ $(\mu =0,1,2,3)$, which
determine the mean-field propagation of the hadrons, and by the
transition rates $G^\dagger G\,\delta^4_\Gamma (\ldots )$ in the collision
term that describe the scattering and hadron production/absorption
rates. For the latter we employ free cross sections as in Ref. \cite{lan91}
that are parameterized in line with corresponding experimental data. In the
relativistic transport approach we explicitly propagate nucleons and 
$\Delta$'s with their isospin degrees of freedom.
For more details we refer the reader to Ref.~\cite{koc91}.

Before going over to a discussion of the scalar and vector self energies
we start with the cascade limit $U^S = U^\nu = 0$ and compare the flow $F$
defined by
\begin{equation} \label{flow}
F = \frac{d}{d \bar{y}} \langle P_x(\bar{y}) \rangle_{\bar{y} = 0}\quad,
\end{equation}
where $\bar{y}$ is the nucleon rapidity in the center-of-mass system normalized
by the projectile rapidity, i.e. $\bar{y} = y_{cm}/y_{proj}$.  
Fig.~1 shows the calculated flow $F$ (divided 
by $A_1^{1/3} + A_2^{1/3})$ as a function of beam energy  
for $Ni + Ni$ and $Au + Au$ systems. We first observe that 
in the cascade mode we obtain a scaling of the flow $F$ with the system size 
expressed by $A_1^{1/3} + A_2^{1/3}$ for both the BUU \cite{hom96} and RBUU
model, which will be solely used later on. It can be observed that the two 
models agree very well below $1 AGeV$, whereas they differ in the energy range
above. 
This difference comes mainly from the fact that the BUU 
includes all higher resonances up to a mass of $2 GeV$ whereas
the RBUU treats only the $\Delta(1232)$ explicitely.
These high mass resonances, though only present in a 
small percentage of all particles in the high density phase, 
contribute significantly to the flow at high energies.
In order to correct the RBUU results for this effect 
when including potentials we add the difference between 
BUU and RBUU in the cascade mode to the RBUU results. 
This correction of course neglects 
higher order effects caused by the smaller velocity of a
heavier resonance in the surrounding nuclear matter; nevertheless, we
expect these corrections to be small.

The flow generated without any mean fields is due
to baryon-baryon collisions and underestimates the experimental data
for $Ni + Ni$ at all energies considerably as shown in Fig.~1. The
difference between the cascade results and the data thus is caused by
the baryon self energies which we try to simulate in the following.

In order to compare our flow values with other theoretical groups, we display 
the quantity
\begin{equation}
\langle P_{x}/N \rangle^{dir} = \frac{1}{N} \int\limits_{-y_{cm}}^{y_{cm}} 
\!\!dy  \,\,
\langle p_x/N\rangle (y) \,\,
\frac{dN}{dy} \,\,\rm{sgn}(y)
\end{equation}
versus beam energy in Fig.~2 for $Au + Au$ systems.
The UrQMD calculations of the Frankfurt group predict the flow for $Au + Au$
collisions to increase with bombarding energy at least up to $4 AGeV$
\cite{uqm96} while hydrodynamical calculations including only hadronic
matter predict a decrease of the baryon flow above  $4-5 AGeV$ 
\cite{ris96} for the heavy system $Au + Au$. Furthermore, 
hydrodynamical calculations 
including a transition to a quark-gluon-plasma (QGP) phase predict 
a vanishing baryon flow at $4-5 AGeV$ \cite{ris96}.
In Fig.~2 we also show the results of RBUU (hard), 
BUU (hard) and UrQMD (hard) EOS calculations as well as our
BUU cascade calculation in comparison with a UrQMD cascade calculation. 
The assignment 'hard' here corresponds to a nuclear incompressibility
$K \approx$ 380 MeV. 
We see, first, that all transport models agree quite well when using a 
'normal' hard equation of state as well as in the pure cascade mode.
All models predict an increasing baryon flow with bombarding energy.
Second, the calculation with a hydrodynamical model shows an even
stronger flow at the energies considered here. 
Third, the RBUU results with a special
momentum-dependence of the mean fields (NL3$^*$, see below)
show a decreasing flow above $1 AGeV$.
This specific model will be discussed in more detail later.

In Fig.~3 we plot the flow $F$ (divided by $A_1^{1/3}+A_2^{1/3}$) 
as a function of the beam energy for both
hard and soft EOS without explicit momentum dependence of the potentials.
The soft EOS is taken from the work of Furnstahl et al.\cite{fur96}
with a nuclear incompressibility $K \sim 194$ MeV while the hard EOS is
the NL3 parameter set from \cite{lan91} with a nuclear 
incompressibility $K=380$ MeV. We notice from this figure that the baryon 
flow in the RBUU approach does not sensitively depend on the nuclear 
incompressibility.
Therefore, in the following we focus on the momentum dependence of these 
potentials only.

In another RBUU approach \cite{li97,sah97} calculations have been done recently
for heavy-ion collisions by employing 
density dependent scalar and vector potentials self-consistently 
as well as the momentum dependence of these potentials while taking 
care of chiral symmetry constraints \cite{fur96}. 
In these approaches the flow has been calculated also for $Ni + Ni$ and $Au + Au$ 
systems and been found to overestimate the experimental data at 1.5 and 
2$AGeV$ for $Ni + Ni$ considerably. Thus also the latter transport calculations 
yield an increasing (or at least constant) flow with bombarding energy, 
whereas the experimental data indicate a decrease for $Ni + Ni$ above $1 AGeV$. 
We will argue that this decrease of the flow $F$ puts stringent constraints
on the momentum- and density dependence of the mean fields.

The model inputs for the mean fields are related to the nuclear 
incompressibility $K$ at density 
$\rho_0$ as well as to the momentum dependence of the mean 
fields as first pointed out by Gale et al. \cite{gal87} and incorporated 
later on also in relativistic transport models
by several authors \cite{mar94,ehe96,li97}. In the RBUU approach
- due to covariance - the scalar and vector mean fields have to be
explicitly momentum dependent \cite{web93} in order to describe 
properly the Schr\"odinger-equivalent optical potential \cite{ham90} 
defined by
\begin{equation}  \label{pot}
U_{sep} (E_{kin}) = U_s + U_0 + \frac{1}{2M} (U_s^2-U_0^2) + 
\frac{U_0}{M} E_{kin}
\end{equation}
as a function of the nucleon kinetic energy $E_{kin}$ with respect to the nuclear 
matter rest frame. However, above $E_{kin} = 1 GeV$ the Schr\"odinger-equivalent
 optical potential is 
not well known experimentally, such that the flow data from the FOPI 
Collaboration could provide further constraints also on this quantity.

In this work
we use a similar Lagrangian density as proposed by Walecka \cite{ser86} 
for the description of nuclear matter, which has been used in the relativistic 
BUU model before \cite{lan91}. This Lagrangian contains nonlinear 
self-interactions of the scalar field $U(\sigma) = \frac{1}{2} m_\sigma^2 
\sigma^2 + \frac{1}{3} B \sigma^3 + \frac{1}{4} C \sigma^4$ 
where the parameters $m_\sigma, B, C$ are calculated 
by fitting the saturation density,
binding energy, effective nucleon mass as well as the compression modulus at 
nuclear matter density (cf. NL3 parameters set from Table 1 in Ref. \cite{lan91}).
An additional coupling between the vector and scalar fields in the
Lagrangian leads to a relatively soft EOS \cite{fur96,fur93}.
In the present calculation we do not consider this effect, 
first, because our main object is to concentrate 
on the momentum dependence on the vector and scalar fields in comparison
to the flow data, second, because the influence of the
 momentum dependence is much stronger than that of the density or
 incompressibility as pointed out before.

The energy density in  mean-field theory (for nuclear matter) in these models 
can be written as \cite{lan91}
\begin{eqnarray}
\varepsilon(m^*,n_b)
&=&g_v V_0 n_b -\frac{1}{2}m_v^2 V_0^2 
+ \frac{m_{\sigma}^2}{2 g_s^2}(m-m^*)^2 
+ \frac{B}{3 g_s^3}(m-m^*)^3     \nonumber \\
&+& \frac{C}{4 g_s^4}(m-m^*)^4 + \gamma\int^{k_f}_{0} \frac{d^3p}{(2\pi)^3}
\sqrt{(p^2+m^*)},  
\label{energy}
\end{eqnarray}
where $m^*=m-g_s\sigma$ is the effective nucleon mass, $n_b$ is
the baryon density and the spin and isospin degeneracy is $\gamma=4$.
$\sigma$ and $V_0$ are the scalar and vector fields with mass $m_{\sigma}$ and
$m_{v}$, which couple to nucleons with coupling constants $g_s$ and $g_v$, 
respectively. The quantities $B$ and $C$ are constant parameters 
and $p$ is the nucleon momentum which has to be integrated up to the 
fermi momentum $k_f$.
In this model the vector and scalar potentials are density dependent, 
however, the vector potential increases only linearly with density. 
Recently \cite{sah97} this Lagrangian has been extended -  maintaining 
chiral symmetry constraints - to include a nonlinear dependence of the
vector potential with density, too. We found, however, that this type 
of Lagrangian density with momentum dependent fields underestimates the 
flow data at all beam energies for $Ni+Ni$ as well as $Au+Au$ systems. 
The calculated flow values follow the cascade results as the scalar 
and vector mean fields cancel each other approximately for such type of 
EOS (cf. Fig.~1).

In our present calculation, we use the energy density Eq.(\ref{energy}) 
for calculating the scalar and vector potentials as a function of density. 
Momentum dependent 
potentials are obtained by fitting the Schr\"odinger equivalent potential 
(\ref{pot}) according to Dirac phenomenology for intermediate energy 
proton-nucleus 
scattering \cite{ham90}. We use the approach from Ref. 
\cite{ehe96} to take into account this momentum dependence by
introducing additional  scalar and vector cutoffs $\Lambda_s$, $\Lambda_v$. 
The scalar and vector form factors at the vertices 
are 
\begin{equation}
\frac{\Lambda_s^2-(p-\langle p\rangle)^2}{\Lambda_s^2+(p-\langle p\rangle)^2} 
\qquad\mbox{and}\qquad
\frac{\Lambda_v^2-(p-\langle p\rangle)^2}{\Lambda_v^2+(p-\langle p\rangle)^2} \quad ,
\end{equation}
respectively, 
where $(p-\langle p\rangle)$ accounts for the difference of the 
one-particle momentum to the average momentum of the surrounding 
nuclear matter.
The values of $\Lambda_s$ and $\Lambda_v$
vary from $0.95$ to $1.05$ GeV and $0.9$ to $1.0$ GeV
to get a good fit to the data. This momentum dependence is not computed 
self-consistently on the mean field level
since it leads only to a small change in the
original parameters of the model as well as in the fitting of the Schr\"odinger 
equivalent potential. So this approximation does not effect much our 
flow results, because for nuclear matter the energy scale involved is 
much smaller ($k_F \ll \Lambda_s, \Lambda_v$) than
in the initial stage of high-energy heavy-ion collisions.

The Schr\"odinger equivalent potential (\ref{pot})
is shown in Fig.~4 as a function of the nucleon kinetic energy with respect to
the nuclear matter at rest;
also plotted are the data from Hama et al. \cite{ham90}.
The solid line in Fig.~4 is for the special momentum dependence including 
form factors
$\frac{\Lambda_s^2-(p-\langle p\rangle)^2}{\Lambda_s^2+(p-\langle p\rangle)^2}$ and
$\frac{\Lambda_v^2-(p-\langle p\rangle)^2}{\Lambda_v^2+(p-\langle p\rangle)^2}$.
The dashed line is obtained using the scalar and vector form factors 
$\frac{\Lambda_s^2}{\Lambda_s^2+p^2}$ and
$\frac{\Lambda_v^2}{\Lambda_v^2+p^2}$
with $\Lambda_s = 0.95$ GeV and $\Lambda_v = 0.9$ GeV, respectively.

\section{Comparison to experimental data}
We use the same parameter sets as for the Schr\"odinger 
equivalent potential in our
flow calculations. The calculations are performed for
the impact parameter $b=4 fm$ for $Ni+Ni$ and $b=6 fm$ 
for $Au+Au$ systems, since for these impact parameters we get the maximum flow,
which corresponds to the multiplicity bins $M3$ and $M4$ as defined 
by the Plastic Ball collaboration \cite{dos87}. 
We have calculated the flow, i.e. the slope parameter (\ref{flow}) by fitting a
straight line from  -0.3$< \bar{y} <$ 0.3 for $Ni+Ni$ and  $Au+Au$ systems 
at all energies. Higher order terms (e.g. fitting a polynomial of 3rd grade)
didn't change the results systematically.

In Fig.~5 and 6 the flow $F$ 
(divided by ($A_1^{1/3}+A_2^{1/3}$) as in Fig.~1) 
is displayed in comparison with the data from Refs. \cite{her96,rei97}
for different systems. Fig.~5 shows the flow for $Ni+Ni$ in the 
energy range up to $2.5 AGeV$ and Fig.~6 for $Au+Au$, respectively. 
In both Figs.~5 and 6 the solid line (NL3) is obtained without 
explicit momentum 
dependence of the self energies, whereas the dashed line
corresponds to the momentum dependent scalar and vector potentials and the
dashed-dotted line (NL3$^{*}$) corresponds to the special 
momentum dependence shown in Fig.~4. 

From Fig.~5 we observe that the dashed-dotted line (NL3$^{*}$) 
is in good agreement with the flow data for $Ni+Ni$, 
whereas for $Au+Au$ (Fig.~6) the 
dashed-dotted line (NL3$^{*}$) is slightly below the flow data from 
\cite{her96,rei97} but closer to the plastic ball data (open triangles).

The important point to be noted here is that the flow rises up to $1 AGeV$ and 
decreases above $1 AGeV$ for $Ni+Ni$, whereas it saturates above $1 AGeV$ for 
$Au+Au$. The physics behind is that the repulsive force due to the vector
mean fields must decrease considerably at high beam energy such that the
Lorentz force on the particles vanishes in the initial phase of the collision.
In subsequent collisions, which are more important in the $Au+Au$ case 
due to its size, the kinetic energy of the particles moving relative to
the local rest frame is then in a range where the Schr\"odinger equivalent 
potential is determined by \cite{ham90}.
We thus conclude that to explain the flow data up to $2 AGeV$ one has
to reduce the vector mean field considerably. 
In other words, there is only a 
weak repulsive force at high relative momenta and high densities.

Another interesting point is that the flow in our calculation for
$Ni+Ni$ decreases earlier with beam energy than for the
$Au+Au$ system. This is due to fact that the flow is governed by the average
transverse pressure generated due to the number of nucleons. For these  
reasons the flow for $Au+Au$ first saturates and then decreases at a higher beam
energy ($\ge 2.5 AGeV$). This implies that the observed
$A^{1/3}$ scaling of the flow up to $1 AGeV$ does no longer hold for 
higher energies. 
Here it would be very interesting to have - besides the $Ni+Ni$ system 
recently measured at FOPI - systematic data for higher mass systems and
beam energies above $1 AGeV$. For $Au + Au$ first preliminary results have been 
presented by the EOS Collaboration \cite{EOS} indicating a gradual decrease 
of the flow $F$ from 2 - 8$AGeV$.

Finally, we show in Fig.~7 the EOS as well as the scalar and vector potential
energy associated
with the special momentum dependence (NL3$^*$). The upper
part shows the energy per nucleon in comparison to the standard NL3
parametrization \cite{lan91} (dashed line). 
For NL3$^*$ (solid line) the approximate equation of state 
has been derived by fitting the vector potential with an appropriate
polynomial in the baryon density. 
The lower part shows the corresponding vector and scalar potential energy
as a function of the baryon density. The vector part for NL3$^*$ is 
substantially lower at high baryon density as compared to
the NL3 parameter set as well as to the original 
$\sigma-\omega$-model \cite{ser86}. The vector potential at $3\rho_0$
is about 460, 653 and 1020$MeV$ for NL3$^*$, NL3 and the 
$\sigma-\omega$-model, respectively.

\section{Summary}
In this work we have calculated the baryon flow in the energy range up to
$2.5 AGeV$ in a relativistic transport model for $Ni + Ni$ and
$Au + Au$. We found that in order to properly describe the flow data of the 
FOPI Collaboration at high beam energies, the strength of the vector potential 
has to be reduced considerably in the RBUU model at high relative momenta
and/or densities. Otherwise, too much flow is generated in the early 
stages of the reaction and cannot be reduced at later phases where the
Schr\"odinger equivalent potential is experimentally known and constrained.

This assumed decrease of the vector and scalar potentials destroys, however,
the $A^{1/3}$ scaling of the flow - 
observed in heavy ion collisions below $1AGeV$ - 
at higher energies. 
Therefore it would be interesting to have systematic studies for 
different mass systems for higher bombarding energies. From these studies 
it might 
even be possible to extract the dependence of the potentials on density and 
momentum separately.

One shortcoming of the transport model used here is the restriction 
to binary nucleon-nucleon or meson-nucleon scattering. 
Especially in meson-nucleon reactions at $2 AGeV$ bombarding
energy the baryon excitations become very high ($\sqrt{s}\approx 2.5GeV$)
such that multi-particle final states can occur which may
lead to a decrease of directed flow.
Despite of these uncertainties above $\approx 2 AGeV$, the decrease
of the Schr\"odinger equivalent potential above $1 AGeV$ 
should have clearly been demonstrated.

\vspace{1cm}
We are happy to acknowledge valuable discussion with G. E. Brown and G. Q. Li. 
One of us (PKS) would like to acknowledge the support from the 
Alexander-von-Humboldt foundation 
and the hospitality of the Institut f\"ur Theoretische Physik at the 
University of Giessen.

\newpage
{\noindent
\small {{\bf Fig.1} The flow $F(y)$ versus beam energy per nucleon for $Ni+Ni$ and
impact parameter $b= 4$ fm and $Au+Au$ systems with impact parameter $b=6$ fm 
in the cascade mode. The solid line corresponds to the RBUU results
for $Au+Au$, the dashed line to the RBUU results for $Ni+Ni$, 
whereas the dotted and dashed-dotted lines correspond to the
BUU results for $Au+Au$ and $Ni+Ni$ systems, respectively.
The experimental data for $Ni+Ni$ are from Ref.\cite{her96}. }}
\vskip 0.3in
{\noindent
\small {{\bf Fig.2}
$\langle P_x/N \rangle^{dir}(y)$ values versus beam energy per nucleon for 
$Au+Au$ at $b=4$fm. A hydrodynamical calculation (for $b=3$ fm) is shown versus 
UrQMD (hard EOS), BUU (hard momentum dependent EOS) and RBUU with momentum 
dependence (NL3) according to \cite{ehe96}.
Also shown are the UrQMD cascade calculations from Ref.\cite{ris96} and the 
BUU cascade calculations \cite{hom96}, which are found to agree very well.
The dashed line results from a RBUU calculation with special momentum 
dependence (NL3$^*$).
}}
\vskip 0.15in
{\noindent
\small {{\bf Fig.3} The flow $F(y)$ versus beam energy per nucleon for $Ni+Ni$ 
at $b=4$ fm.
The dashed line results from the soft EOS from Ref. \protect\cite{fur96} while the 
solid line is obtained for a hard EOS \protect\cite{lan91}. 
The data points are from the FOPI Collaboration 
\protect\cite{her96}.\mbox{}}}
\vskip 0.15in
{\noindent
\small {{\bf Fig.4} The Schr\"odinger equivalent potential (\protect\ref{pot}) as a 
function of the nucleon kinetic energy $E_{kin}$. 
The solid curve is for the special momentum dependence NL3$^*$
discussed in the text
and the dashed curve for the momentum dependent parameter set NL3 
\protect\cite{ehe96} (see text). 
The data points are from Hama et al. \protect\cite{ham90}.\mbox{}
}}
\vskip 0.15in
{\noindent
\small {{\bf Fig.5} The flow $F(y)$ versus the beam energy per nucleon for $Ni+Ni$ 
at $b=4$ fm. 
The solid line results for the parameter set NL3, 
the dashed line includes an explicit momentum dependence of NL3 and 
the dashed-dotted line is for the special
momentum dependence NL3$^*$, where the vector potential decreases 
at high energies (c.f. Fig.~4). 
The data points are from the FOPI Collaboration \protect\cite{her96}.\mbox{} }}
\vskip 0.15in
{\noindent
\small {{\bf Fig.6} Same as Fig.5 for $Au+Au$ at $b=6$ fm. The data points 
are from the Plastic Ball, FOPI and EOS Collaborations, 
Ref.\protect\cite{her96}.\mbox{}}}
\vskip 0.15in
{\noindent
\small {{\bf Fig.7} Upper part: The solid line shows the equation of state for
the parameter set NL3$^*$ in comparison to NL3 (dashed line).
Lower part: Vector (upper solid line) and scalar (lower solid line) potential
energy per nucleon for NL3$^*$ in comparison to NL3 (dashed line).
}
\vskip 0.15in

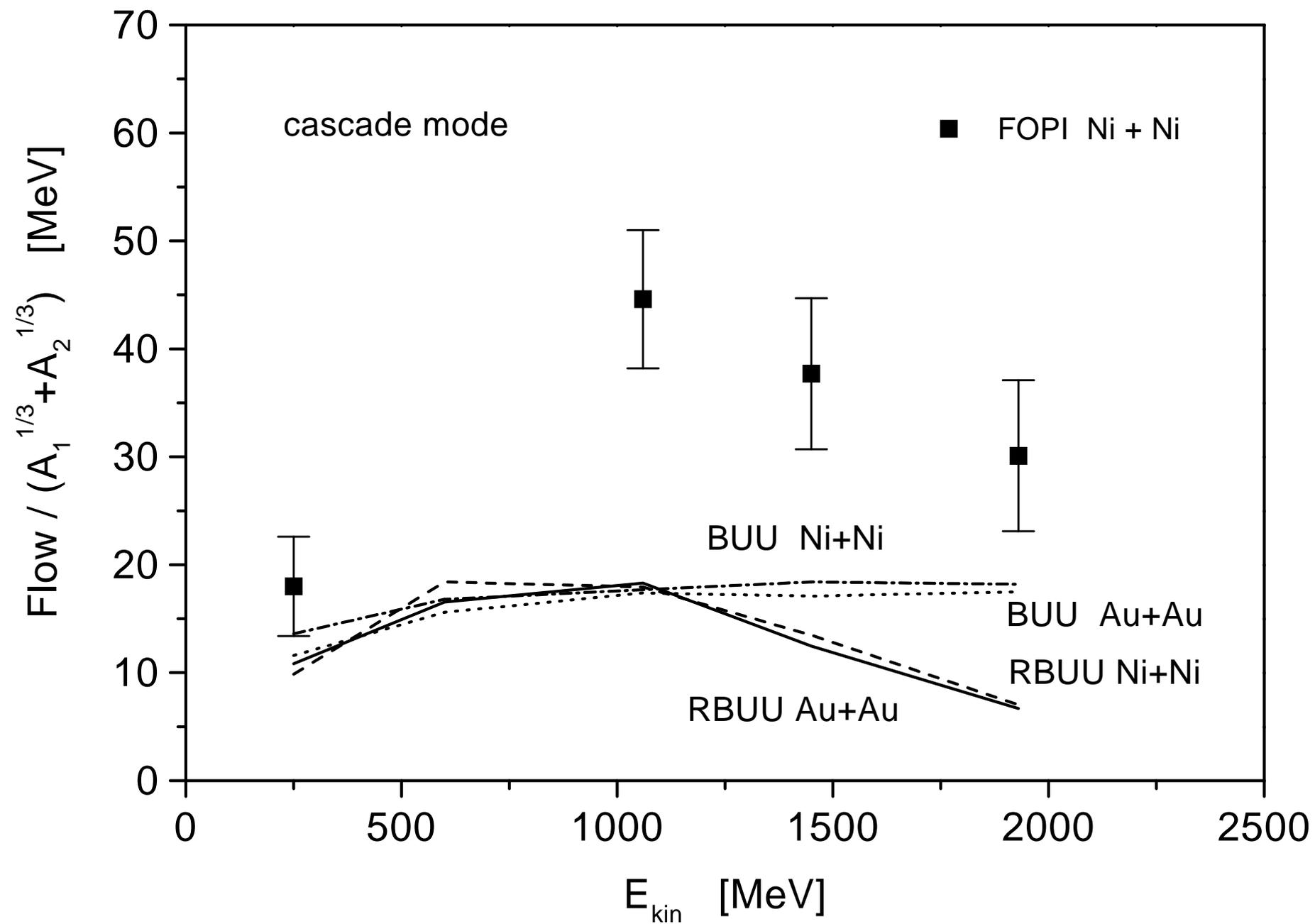
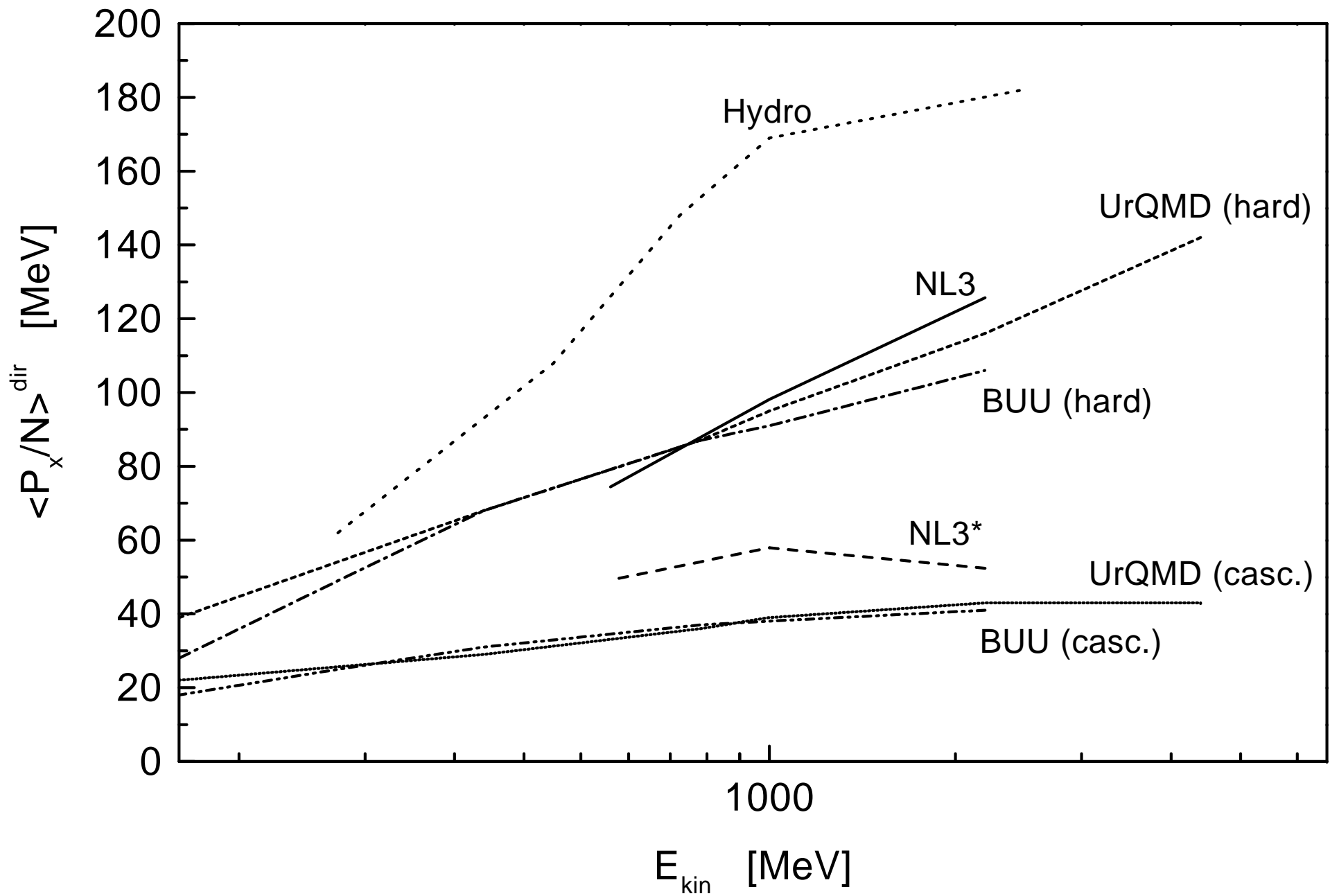
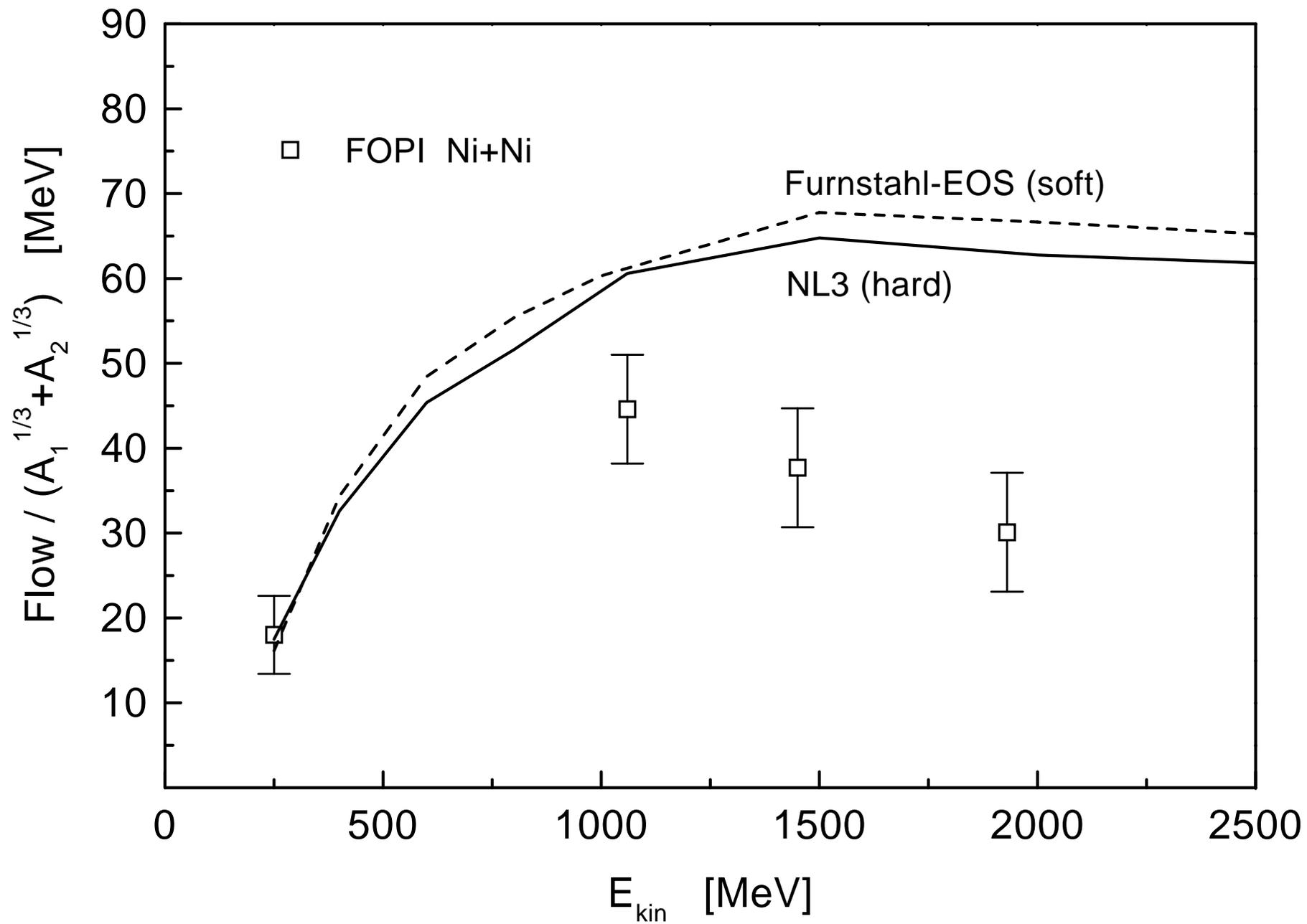
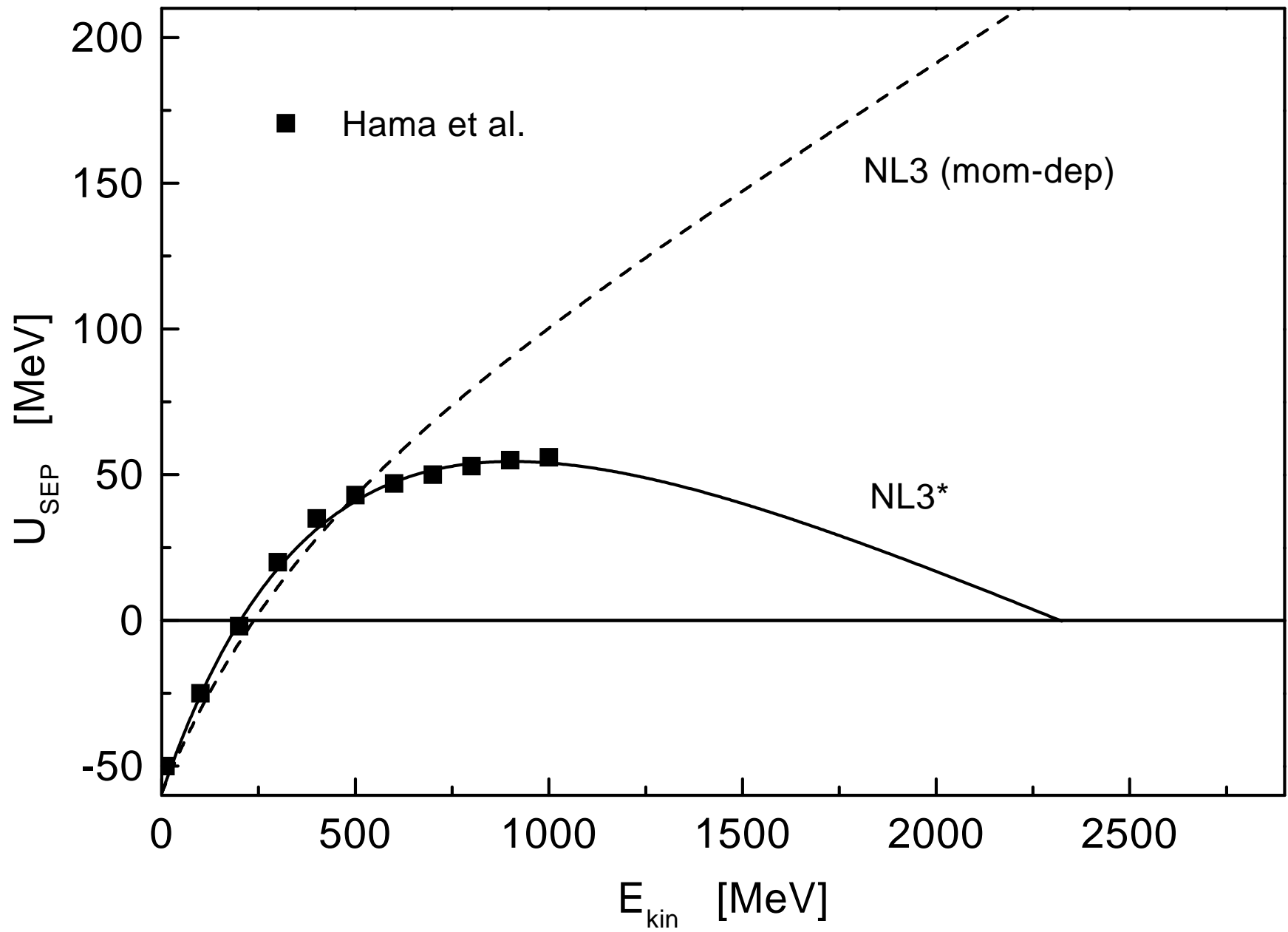
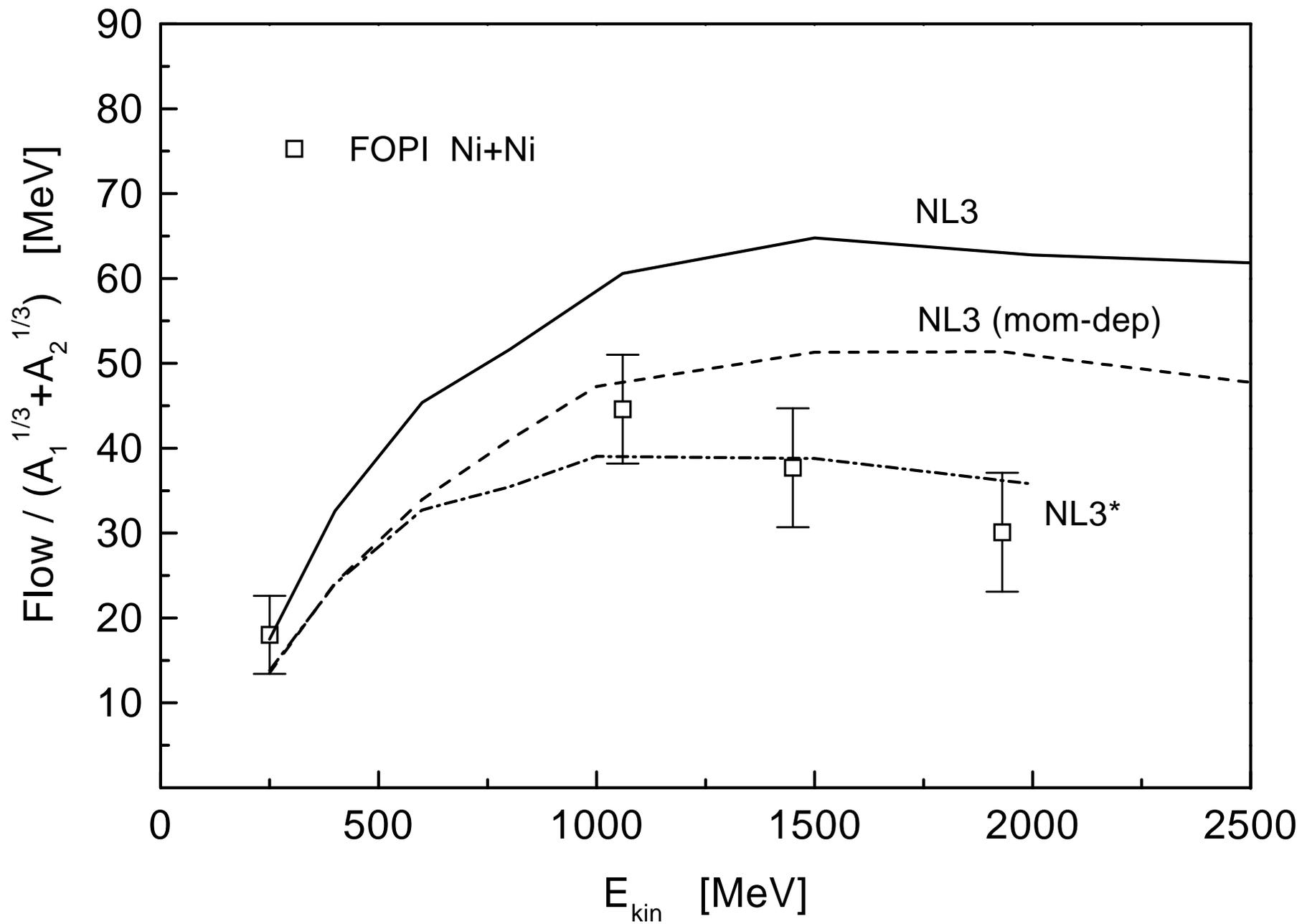
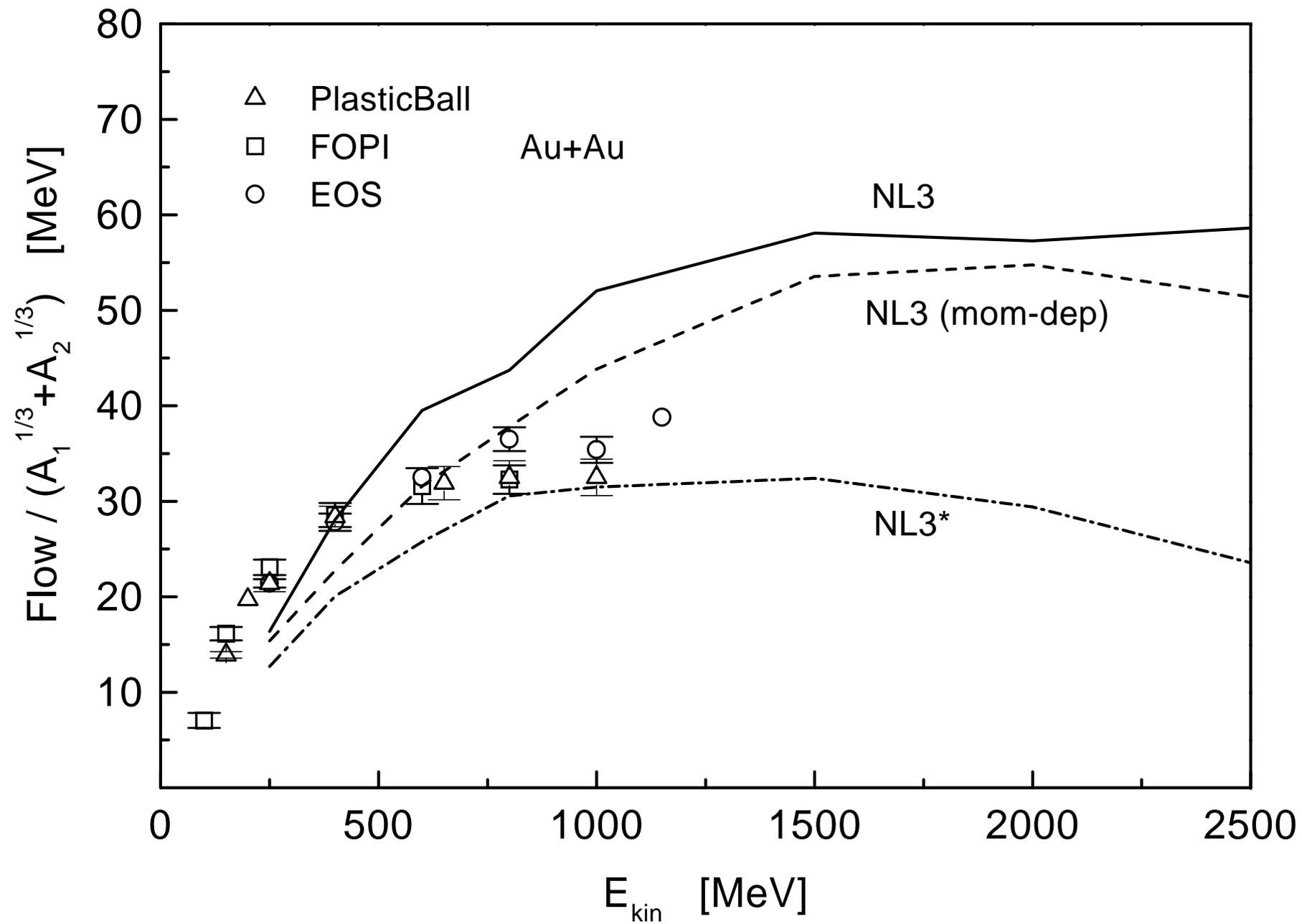
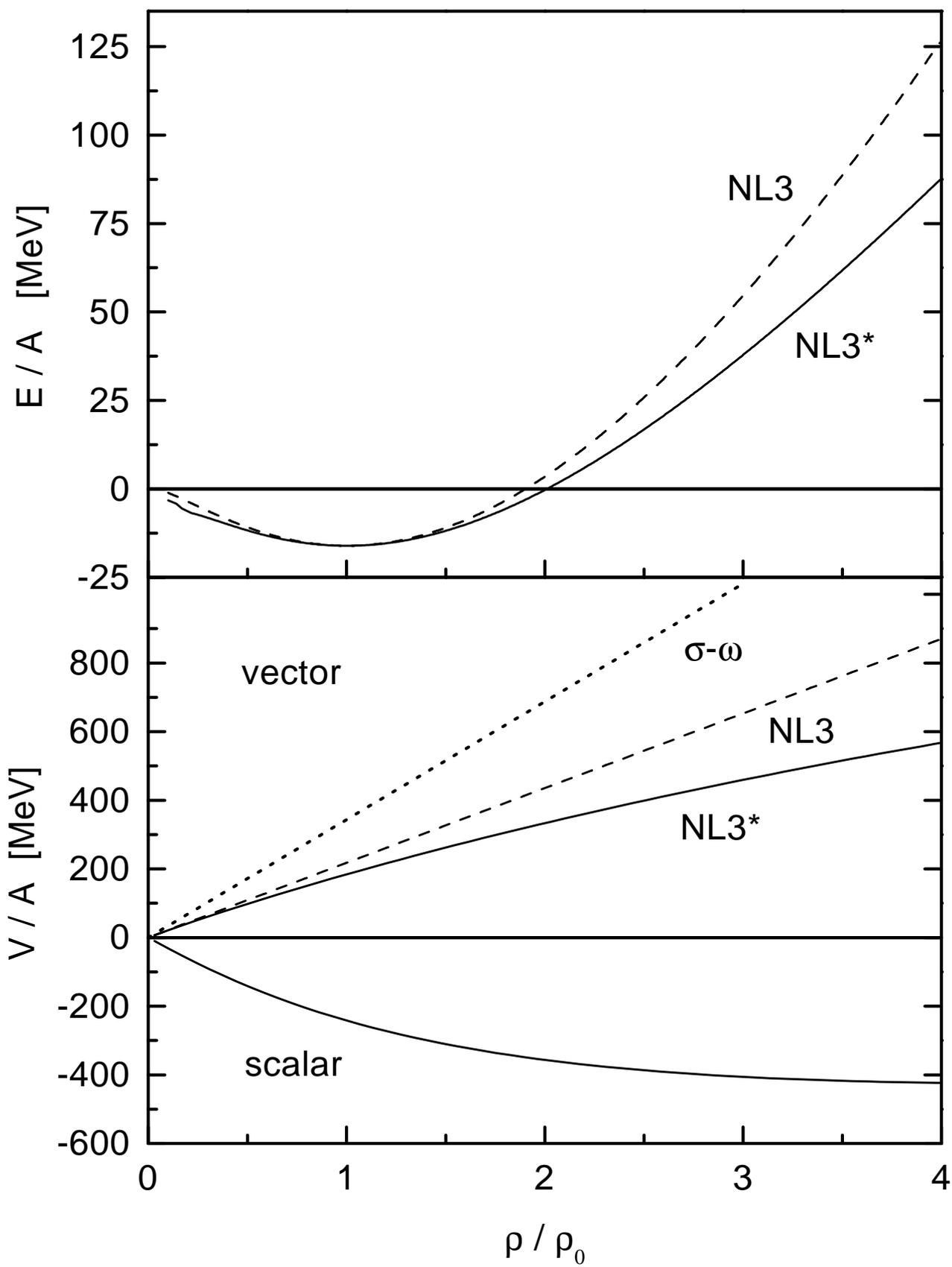
\end{document}